\documentclass[num-refs]{nbdt-article}

\usepackage{siunitx}

\usepackage{amsmath, amsfonts}

\papertype{Original Article}
\paperfield{Journal Section}

\title{Neural network-based encoding in free-viewing fMRI with gaze-aware  models}

\author[1]{Dora Gözükara}
\author[1]{Nasir Ahmad}
\author[2,3]{Katja Seeliger}
\author[1]{Djamari Oetringer}
\author[1]{Linda Geerligs}

\affil [1] {Donders Institute for Brain, Cognition and Behaviour, Radboud University, Nijmegen, the Netherlands}
\affil [2] {Martin Luther University Halle-Wittenberg, Medical Faculty, Halle, Germany}
\affil [3] {Max Planck Institute for Human Cognitive and Brain Sciences, Leipzig, Germany}

\corraddress{Dora Gözükara}
\corremail{dora.gozukara@donders.ru.nl}

\runningauthor{Gözükara et al.}

\begin{document}

\maketitle

\begin{abstract}
\small
Representations learned by convolutional neural networks (CNNs) exhibit a remarkable resemblance to information processing patterns observed in the primate visual system on large neuroimaging datasets collected under diverse, naturalistic visual stimulation, but with instruction for participants to maintain central fixation.
This viewing condition, however, diverges significantly from ecologically valid visual behaviour, suppresses activity in visually active regions, and imposes substantial cognitive load on the viewing task.
We present a modification of the encoding model framework, adapting it for use with naturalistic vision datasets acquired under fully natural viewing conditions, without fixation, by incorporating eye-tracking data.
Our \textsl{gaze-aware encoding models} were trained on the StudyForrest dataset, which features task-free naturalistic movie viewing.
By combining eye-tracking data with the visual content of movie frames, we generate combined subject-wise gaze-stimulus specific feature time series.
These time series are constructed by sampling only the locally and temporally relevant elements of the CNN feature map for each fixation.
Our results demonstrate that \textsl{gaze-aware encoding models} match the performance of conventional encoding models with 112$\times$ fewer model parameters.
Gaze-aware encoding models were especially beneficial for participants with more dynamic eye-movement patterns.
Therefore, this approach opens the door to more ecologically valid models that can be built in more naturalistic settings, such as playing games or navigating virtual environments. 

\end{abstract}

\section{Introduction}
Cognitive neuroscience is increasingly adopting integrative research paradigms that combine naturalistic experimental designs with advanced analytical tools.
For instance, the use of complex, multimodal and naturalistic stimuli in experiments, paired with the application of artificial neural networks (ANNs), has gained traction in recent years.
Representations learned by convolutional neural networks (CNNs), in particular, have become a prominent model for understanding stages of information processing in the primate visual system \citep{neuroconnectionism, lindsay2021cnnvis}.
These developments align with a broader effort to make experimental designs and analytical methods more ecologically valid.
This paper builds on these advances by revisiting assumptions underlying brain-encoding models that rely on CNN features.
Specifically, we propose that incorporating natural eye movements into the modelling process can significantly enhance the ecological validity of the underlying data and reduce the parameter spaces of models.
By integrating gaze dynamics we demonstrate that CNN-based brain encoding can accommodate naturalistic stimuli without the need for restrictive fixation in the experimental designs.

\subsection{Naturalistic Experimental Design }
Natural visual behaviour is inherently and consistently active, with frequent gaze shifts to explore salient features in the environment \citep{hayhoe2011vision}.
Despite this, almost all experiments underlying recent deep neural network (DNN)-based brain modelling studies have imposed central fixation constraints during neuroimaging experiments to control attentional focus and minimize variability introduced by eye movements \citep{kay2008identifying, nishimoto2011vim2, horikawa2017generic, allen2022NSD, chang2019bold5000, hebart2023things}.
While this approach ensures stimulus-response measurements that are simple to handle analytically, it diverges from natural viewing behaviour and imposes substantial cognitive load.

Fixation protocols also suppress activity in visually dynamic brain regions, potentially skewing insights into how the brain processes naturalistic stimuli.
For example, \citet{dorr_variability_2010} highlight that fixating on a static location during movie viewing disrupts the natural flow of perception.
Nevertheless, some large neuroimaging data collections required participants to maintain fixation for extended periods, up to several dozens of hours \citep{seeliger_large_2021, allen2022NSD}.
Even in studies incorporating eye-tracking data, its use is typically limited to verifying fixation compliance or assessing participant alertness \citep{kauttonen_brain_2018, zadbood_how_2017}.

Here we propose to use eye-tracking data to inform feature selection in CNN-based brain-encoding models.
This approach allows for fixation-free experiments that better capture the dynamic and active nature of visual behaviour, offering a path toward more ecologically valid experimental designs and lowered computational burdens.
This research is particularly relevant since some of the largest ongoing neuroimaging data recording efforts have decided against fixation protocols for aforementioned reasons \citep{boyle2020courtois}.

\subsection{Holistic Analysis }
Advances in artificial neural networks, particularly CNNs, have revolutionized our ability to model brain activity.
Early brain-encoding models relied on manually engineered features or simple statistical representations \citep{kay2008identifying, nishimoto2011vim2}.
By contrast, CNNs learn hierarchical visual representations directly from large-scale naturalistic data, making them a powerful tool for studying information processing in the visual system \citep{kriegeskorte2015deep, lindsay2021cnnvis, neuroconnectionism}.
One key similarity between CNNs and brains lies in their use of spatially selective receptive fields, where individual neurons or units respond to specific regions of the visual field.

Most fMRI studies using CNN-based encoding models do not leverage these spatial properties and pool features from entire CNN layers into single feature vectors to predict voxel activity \citep{agrawalCNN, agrawal2014pixels, guclu_deep_2015, horikawa2017generic, wen_neural_2018, han_variational_2019,Kamali2025}.
Limited exceptions exist, such as recent works which transform input images to mimic cortical magnification of the visual input \cite{daCosta2024}.

The use of features gathered from across all spatial locations in CNN layers inflates model parameter spaces, requiring large datasets to achieve reliable fits and introducing ambiguities in feature selection, given that there are likely multiple combinations of features and locations that could explain the signal of a voxel.
By sampling CNN features with eye-movements instead, we propose a significant reduction in parameter space complexity, improving both data efficiency and model interpretability.

\subsection{Gaze-Aware Brain Encoding }
The primary goal of this work is to demonstrate the efficacy of combining the power of convolutional neural networks (CNNs) and eye-movement data to model the brain's visual processing system in experimental conditions that are closer to natural viewing scenarios.
Specifically, in experimental neuroimaging scenarios that allow for eye movements.
We further contribute to modelling efficiency by introducing a method that modifies brain-encoding models to significantly reduce the parameter space and thus the amount of data required for training.
To demonstrate this, we used the publicly available StudyForrest dataset \citep{hanke_studyforrest_2016}, which includes approximately two hours of fixation-free movie-watching runs and comprehensive eye-tracking data during the movies.
Using a CNN pre-trained for natural image classification, we extracted features from all convolutional layers in response to the movie frames.
This generated a time series of features for each spatial location and layer corresponding to each movie frame. 
We sampled these features using subject-specific gaze data to construct personalized feature time series.
The resulting feature time series were highly efficient in size, reducing feature data size by factors ranging from 112-fold for the smallest layer to approximately 30,000-fold for the largest layer.
Finally, we trained linear encoding models to learn optimal feature combinations from these reduced temporal feature series that predict voxel activity.
By integrating gaze dynamics, this pipeline not only addresses the variability across and within individuals but could also enable robust predictions using substantially smaller datasets.
This represents a meaningful step forward in making neural encoding models more efficient and practical for real-world applications, particularly in scenarios where traditional experimental constraints, such as fixation, are relaxed. 

\section{Methods}

\subsection{fMRI Dataset }
We make use of the open ‘StudyForrest’ dataset which, alongside the taskless movie viewing sessions (of the movie Forrest Gump dubbed in German, presented across 8 runs of around 15 minutes each), includes retinotopic mapping runs as well as eye-tracking for 15 subjects (aged 19 to 30).
Further details about the acquisition can be found in the original publications \citep{hanke_studyforrest_2016,sengupta_studyforrest_2016}.

\subsubsection{Movie Viewing Runs }
The fMRI dataset was provided both preprocessed and denoised by \citet{Liu2019}.
The pre-processing steps are described in detail in their paper and are briefly summarized here. 
Their pre-processing steps included motion correction, slice timing correction, brain extraction, and high-pass temporal filtering with a 200 s cut-off.
They also used spatial ICA decomposition to remove head motion, physiology, and hardware related artefacts.
We further pre-processed the data by carrying out co-registration with each subject's T1 anatomical image, followed by segmentation and normalisation to Montreal Neurological Institute (MNI) space. We used SPM 12 for all pre-processing steps above (\href{http://www.fil.ion.ucl.ac.uk/spm}{www.fil.ion.ucl.ac.uk/spm}).
After anatomical normalisation, each voxel's timecourse within each run was z-scored independently.
Finally, we concatenated the 8 runs resulting in a single functional timeseries.
We limited our analyses to brain regions that were expected to encode visual information, based on the AAL atlas \citep{tzourio-mazoyer_automated_2002}.
This includes areas in the ventral visual stream, parts of the dorsal stream and the temporal lobe.

\subsection{Eye-Tracking }\label{eyetrack}
Eye tracking was performed using monocular corneal reflection and pupil tracking with an Eyelink 1000 (software version 4.594) equipped with an MR-compatible telephoto lens and illumination kit.
More information about the details can be found in the original paper \citep{hanke_studyforrest_2016}.
To get a list of all fixation events per participant, and their coordinates on the screen, we used Remodnav \citep{dar_remodnav_2020}.
Remodnav is an algorithm for classifying eye-tracking events, which is founded on a pre-existing velocity-oriented approach, and can categorise saccades, post-saccadic oscillations, fixations, and smooth pursuit occurrences.
For simplicity, we only used fixation events and ignored saccades or pursuits in our models.
We used the default parameters of Remodnav to define our fixation events.
We then matched the fixation event timestamps to the movie frames. 
In addition to determining the spatial location of our CNN features, we also used the fixation events to select which movie frames to include in our gaze-aware encoders.
Instead of sampling the same equidistant frame set for all participants, we sampled the frames based on when each participant fixated.
As fixation events span multiple frames, we only used the middle frames of each fixation event to extract the CCN features for the encoding model.
Because of the low quality of the eye-tracking data of 2 participants, we excluded them from our study, leaving us with 13 participants.

\subsection{CNN Features }
Our proposed approach can in principle be used with any artificial neural network model which has accessible spatially variant features.
In this work, we used a pre-trained CNN network but alternative architectures such as image transformers, generative adversarial networks, or diffusion networks could also be used. 

\textbf{Model:}
The CNN model we used to extract movie frame features is the Pytorch implementation of VGG-19, pre-trained to classify ImageNet images \citep{simonyan_very_2015}.
We used the responses of 5 max pooling layers, each of which following a block of convolutional layers, to build our encoding models.
Fully connected layers were removed from the architecture as they lack spatial structure and have no spatial dimension.
Furthermore, the fully connected layers of the model require images with a fixed 1:1 aspect ratio.
Therefore, removing the fully connected layers made it possible to use the custom (cinematic) aspect ratio of 2.35:1 of the movie frames.

\textbf{Input Images:}
To convert the movie stimulus into images that could be used as input to VGG-19, we cropped the grey bars above and below the movie frames, and downsampled them while respecting the aspect ratio (original (1280, 544), resized (527, 224)).
We then applied the default VGG input image normalisation to the pixel values for each RGB channel with means of 0.485, 0.456, 0.406, and standard deviations of 0.229, 0.224, 0.225 respectively. 

\textbf{Feature Maps:}
After getting a list of all fixation events as described above, we obtained feature maps from our CNN model for each frame corresponding to the timepoint of fixation.
The five feature maps are extracted from different layers of the CNN model, each with differing spatial and channel dimensions.
We temporally binned these features by averaging all frame features (or gaze-aware feature) that fall within a TR of the fMRI measurements. 
For computational feasibility, and to be able to combine features across different layers in our models, we fixed the spatial sizes of feature maps of all layers to the same size (7,16) by doing a spatial rescaling through averaging of features.
All of these spatially rescaled feature maps were then concatenated to form a ‘hyperlayer’ feature map.
This process was a pragmatic choice, and gaze-aware encoding models may also be built for individual layers with their native resolutions.

\begin{figure*}[ht]
\begin{center}
\includegraphics[width=\textwidth]{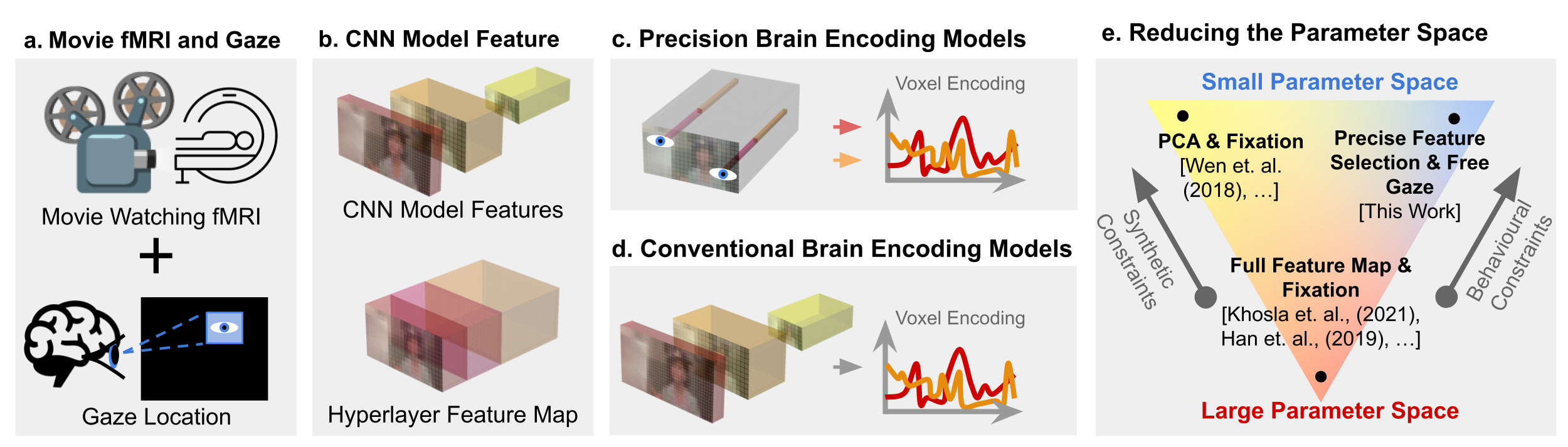}
\end{center}
\caption{\textbf{Schematic figure that outlines the most important aspects of gaze-aware encoding models.}
    (\textbf{a}) Unique gaze patterns for each individual are collected. 
    (\textbf{b}) For computational efficiency, and to be able to build one model across all layers, we combined CNN model features into a hyperlayer feature map.
    However, gaze-aware encoding models could also be built for separate layers.
    (\textbf{c}) This way, the gaze-aware encoding model in each voxel is based specifically on the CNN features that are relevant for information processed by each individual.
    (\textbf{d}) Traditional CNN based encoding models use the whole feature sets of layers.
   (\textbf{e}) Diagram which shows how our approach compares to other approaches in the literature in terms of the number of parameters in the model and the type of constraints that are used to reduce the number of parameters (synthetic or behavioural constraints).} 
\label{MethodsExpo}
\end{figure*}

\subsection{Gaze-Aware Encoding Model }
To create our proposed gaze-aware encoding models, we collected the gaze locations at each frame for every participant.
This allowed us to extract a subject specific feature timeseries by sampling only the parts of the visual scene that the subjects looked at at each timepoint.
This feature timeseries was then used to train a linear encoder that estimates the mapping from features in the timeseries to voxel activity.
Herein we describe the pipeline by which this data is extracted and trained, see Figure \ref{MethodsExpo} for a visualization.

\textbf{Feature map extraction: }
Given $N$ fixation timepoints, corresponding movie frames are extracted to create a dataset of images, $I \in \mathbb{R}^{N \times H \times W \times 3}$ where $N$ is the number of frames, $H$ is the frame height, $W$ the frame width, and each pixel has three values (RGB). 
Each frame is then preprocessed (normalized and downsampled) and then passed through the CNN model (VGG-19) while collecting all activations at max-pooling layers.
This results in a set of feature maps, $A_0, A_1, A_2, A_3, A_4$, each of which has a dimension of the number of frames, as well as independent dimension of height, width, and channel.
In particular, $A_n \in \mathbb{R}^{N \times H_i \times W_i \times C_i}$ where subscripts of $i \in [0, 1, 2, 3, 4]$ indicate specific layer feature map heights ($H_i$), widths ($W_i$), and channels ($C_i$).

Finally, all feature maps are combined to create a "hyperlayer" feature map via a spatial down/up-scaling to a  height and width of  $\bar{H}=7$, and $\bar{W}=16$, respectively. All channel dimensions are concatenated, $\bar{C} = (C_0 + C_1 + C_2 + C_3 + C_4)$
This hyperlayer feature map can be denoted $M \in \mathbb{R}^{N \times \bar{H} \times \bar{W} \times \bar{C}}$, note that channel-wise concatenation simply results in a summed channel dimension with a total number of features $\bar{C} = 1472$.

\textbf{Gaze Adjusted Features: }
Given a unified hyperlayer feature map, $M$, we make use of gaze estimations in order to further refine the data used for the fitting of the encoding model.

Let the gaze coordinates for a given frame, $n$ be represented by $x^\text{gaze}_n$ and $y^\text{gaze}_n$.
These coordinates are assumed to be computed in pixel-space.

Using these as (normalized) coordinates, one can extract a subject specific feature set from the hyperlayer feature map, such that

\begin{equation}
    X^\text{gaze}_{n} = M_{n, x^\text{gaze}_{n}, y^\text{gaze}_{n}, :}
\end{equation}

where $X^\text{gaze}_{n}$ is a frame, gaze-specific feature vector.
From this, the full gaze-aware feature vector can be written $X^\text{gaze} \in \mathbb{R}^{N \times \bar{C}}$, where $X^\text{gaze} = [X^\text{gaze}_{0}, X^\text{gaze}_{1}, X^\text{gaze}_{2} ... X^\text{gaze}_{n}]^\top$.

\textbf{Hemodynamic response function adjustment: }
The feature vectors were averaged over all fixations within a single fMRI sample duration for every voxel to form a full features stack of size $[\text{\#Voxels}, 1472]$, and this feature stack, shifted 4.5 seconds forward in time to account for the hemodynamic response function (HRF), was the final data used to predict voxel activity.

\textbf{Linear encoding model: }
Using the gaze specific set of feature vectors, $X^\text{gaze}$, one may build a linear encoding model:

\begin{equation}
    Y = X^\text{gaze}W
\end{equation}

where $W \in \mathbb{R}^{\bar{C} \times V}$ is the weight matrix, $V$ is the number of voxels being predicted, and the encoding $Y \in \mathbb{R}^{N, V}$ therefore represents the voxel-wise timeseries signal.
The total number of parameters in our gaze-aware weight matrix, $W$, comes to $\text{\#Voxels} \times 1472$.
Our choice of concatenating all feature maps into a hyperlayer allows our encoding model to train a single matrix of parameters, computationally simplifying the setup.
Alternative methods often train an encoding model for each individual layer (feature map) of a CNN, while this method trains only one based upon the hyperlayer.

Finally, the optimal weights $W$ that best describe the relationship between the stimulus $X$ and the neural response $Y$ can be determined via ridge regression using the Pearson correlation as a loss function (1-Pearson r):

\begin{equation}
\mathcal{L}(W) = (1 - \frac{\text{cov}(XW, Y)}{\sigma_{XW}\sigma_Y}) + \lambda||W||_2^2
\end{equation}

\textbf{Determining the Ridge regularization space for each subject:}
We estimated the optimal regularization hyperparameter by searching a grid 10 values spaced logarithmically from 0.1 to 100,000,000.
This range was large enough to ensure that the algorithm explored many different possible regularization strengths for each feature space, and we ensured that the optimal value was never the highest tested.

\subsection{Experiments and Baselines }
Code for reproducing the results in this work is available at \href{https://github.com/drgzkr/GazeAwarefMRIEncoding}{https://github.com/drgzkr/GazeAwarefMRIEncoding}.
For all our baseline models described below, the same approach of averaging features per TR, HRF adjustment, and ridge regression was used as for the gaze-aware model described above. 

\textbf{Baseline model:}
As our main baseline model, we used the entire hyperlayer feature maps, without taking gaze location into account.
The feature timeseries for the baseline model are still subject specific because they were sampled at each participants’ fixation event timepoints (just like in the gaze-aware model).
We then trained a linear encoder that learns the best combination of features in this timeseries that predicts the activity of the voxel.
This baseline linear encoder model has a total number of parameters $\text{\#Voxels} \times 164,864$ ($\bar{H}\times\bar{W}\times\bar{C} = 7 \times 16 \times 1472$), 112 times larger than the gaze-aware model.

\textbf{Center fixation baseline model:}
To isolate the benefit of personalised spatial selection from any reduction in feature dimensionality, we trained a model that constrains spatial sampling to the center of each frame.
Specifically, for each fixation event the feature vector is extracted from the center location of the hyperlayer feature map, yielding a vector of size $[1 \times \bar{C}]$ — identical in dimensionality to the gaze-aware model's per-frame feature vector.
As a result, both models have the same number of parameters ($\text{\#Voxels} \times 1472$); the only difference is \textit{which} spatial location is sampled: the center of the frame versus each participant's actual fixation location.
This ensures that any performance difference between the two models reflects the informativeness of the sampled location, not a difference in model capacity.

\textbf{PCA baseline model:}
To understand if constraining the feature space spatially is meaningful, we included a model which had the same number of features as the personalised models, but this dimensionality reduction was achieved by using the first 1,472 principal components of the full feature space.

\textbf{Memory footprint:} 
A practical advantage of gaze-aware encoding is the reduction in memory required both to store the feature time series and to fit the encoding model.
All quantities below can be derived analytically from the pipeline parameters and are summarised in Table~\ref{tab:memory}.
The feature time series fed to ridge regression forms the design matrix $X \in \mathbb{R}^{N \times F}$, where $N = 3{,}599$ is the number of fMRI volumes and $F$ is the number of features per volume.
For the baseline model, $F = \bar{H} \times \bar{W} \times \bar{C} = 7 \times 16 \times 1{,}472 = 164{,}864$, whereas for the gaze-aware model $F = \bar{C} = 1{,}472$.
At float32 precision (4 bytes), this yields design matrices of 2.37~GB and 21~MB respectively.
The weight matrix $W \in \mathbb{R}^{F \times V}$ (where $V = 19{,}629$ voxels) follows the same ratio: 12.94~GB for the baseline versus 116~MB for the gaze-aware model.
The response matrix $Y \in \mathbb{R}^{N \times V}$, which is identical for both models, contributes an additional 283~MB.
In total, peak working memory during ridge regression is approximately 15.6~GB for the baseline model and 419~MB for the gaze-aware model — a \textbf{37-fold} reduction.
This brings gaze-aware encoding within the memory budget of a standard laptop, whereas the baseline model requires a workstation or HPC allocation.
\begin{table}[ht]
\centering
\footnotesize
\caption{
    \textbf{Memory footprint comparison: baseline vs.\ gaze-aware encoding model.}
    All quantities are derived analytically from the pipeline parameters
    (see Methods). 
    Storage assumes float32 (4 bytes per value).
    Working memory refers to the peak RAM required during ridge regression
    (design matrix $X$ + response matrix $Y$ + weight matrix $W$).
}
\label{tab:memory}
\setlength{\tabcolsep}{25pt} 
\begin{tabular}{lrr}
\hline
\textbf{Quantity} & \textbf{Baseline} & \textbf{Gaze-aware} \\
\hline
Features per TR                     & 164,864 B & 1,472 B \\
Model parameters (\#voxels $\times$ features) & 3,236,115,456 B & 28,893,888 B \\
Design matrix $X$                   & 2.37 GB & 21.19 MB \\
Weight matrix $W$                   & 12.94 GB & 115.58 MB \\
Response matrix $Y$ (shared)        & 282.58 MB & 282.58 MB \\
\hline
\textbf{Total working memory}       & \textbf{15.60 GB} & \textbf{419.35 MB} \\
Reduction factor                    & \multicolumn{2}{c}{$37\times$} \\
\hline
\end{tabular}
\end{table}

\section{Analysis}

\subsection{Training and testing data }
The fMRI dataset has 3599 total volumes (with 2 seconds TRs) divided between 8 runs.
We separated the middle 20 percent of all timepoints (655 volumes in total) from each run to use as our test set.
We further divided the training set into a validation set (20 percent, 574 volumes) and a training set (80 percent, 2306 volumes) to search for the best regularization value for ridge regression.
We also removed 4 TRs from beginning and end of each section to remove the remaining autocorrelation between our test and training sets (64 volumes).
Using the validation and training sets, we trained 5 different encoders using the same set of features but different regularization values and determined the best regularization parameter.
We then trained the encoder with the entire training set using the best regularization value before measuring its accuracy (Pearson correlation) on the test set.

\subsection{Statistical testing and visualization }
As each of our single subject volumes were projected to MNI space, we were able to test the significance of model performance across all 13 participants.
For each model and voxel, we performed a one-sided  paired t-test (alternative hypothesis being the performance is greater than zero) across participants to determine if a particular model can statistically significantly predict that voxel's activity.
We then applied False Discovery Rate \citep{Benjamini1995} correction with the total number of voxels (19629) to correct for multiple testing.
In order to identify which areas are predicted better than the others, we used the Julich atlas version 2.9 \citep{Amunts2021} and computed the average performances of voxels within a region of interest (ROI) for each participant and model. 
We used the following six ROIs: V1 (human Occipital cortex 1 (hOc1)), V2 (hOc2), V3 (hOc3), lateral occipital (LO) areas (made up of hOc4p, hOc4la, and hOc5), fusiform gyri (FG; made up of FG1, FG2, FG3, FG4), and superior temporal sulcus (made up of STS1 and STS2).
For generating the surface maps, we used the volume to surface projection tools of the Python package Nilearn \citep{Nilearn}, by projecting the fMRI data in MNI space to a common mesh. 

\subsection{Noise ceiling omission }
Unlike most fMRI encoding work, our work does not include a noise ceiling estimates. 
This means that our performance estimates cannot be directly compared to those in previous studies.  
Noise ceilings are often estimated using data collected from repeated test stimuli, which our free-viewing dataset does not include.
Note, it is also important to consider that the noise ceiling estimated via a typical repeated (fixed-gaze) test stimuli would not be appropriate for free-viewing naturalistic datasets.
Other metrics like inter-subject synchrony (ISS), which have also be used as an estimate for noise ceiling, were considered but could also distort the differences we observe between brain regions for our models, which rely upon subject specific gaze.
ISS would be more affected by these inter-individual differences especially in lower order areas.

\section{Results}

\begin{figure*}[!ht]
\begin{center}
\includegraphics[scale=0.75,width=0.75\textwidth]{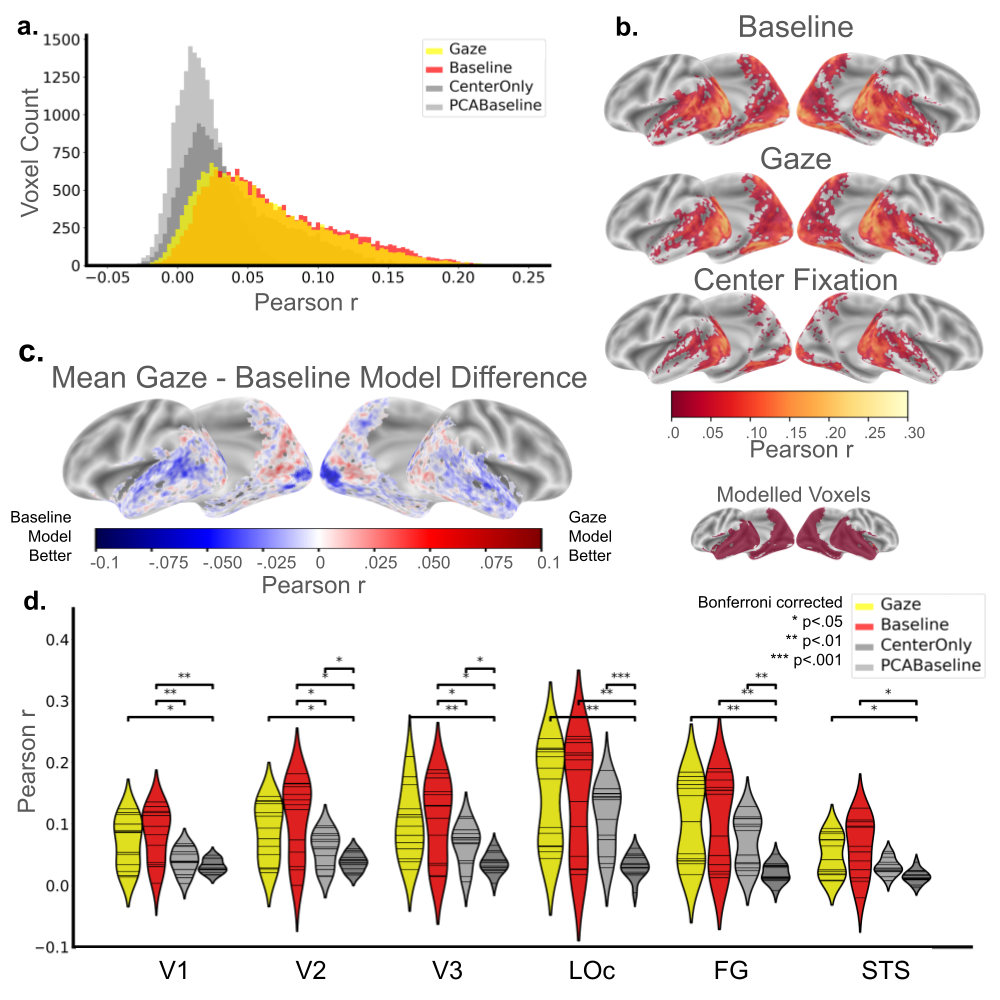}
\end{center}
\caption{\textbf{Gaze-aware models match baseline models.}
    (\textbf{a}) Histograms showing group average model performances for all modelled voxels.
    (\textbf{b}) Cortical maps show group average model performances for voxels that are statistically significantly predicted by each model.
    Gaze-aware and baseline models predict the same range of voxels with very similar performances.
    Center fixation models predict fewer voxels.
    PCA baseline models reach statistical significance for only 3\% of voxels, and thus are not shown.
    (\textbf{c}) Cortical map shows group average model performance differences between the gaze-aware model and baseline model.
    (\textbf{d}) Violin plots show model performances in bilateral V1, V2, V3, LOc, and FG.
    Each violin plot shows the distribution of average model performance within an ROI across all participants.
    Each horizontal line shows one participant.
    Gaze-aware models are not statistically significantly different than baseline models in any ROI.} 
\label{fig:fig1}
\end{figure*}

\subsection{Model performances }
We assessed the performances of different encoding models for each voxel by using a paired-sample t-test across all 13 participants.
We found that gaze-aware encoding models predicted 53\%  of all voxels in our mask statistically significantly after FDR correction, compared to 57\% for baseline models.
Other control models showed much lower performance; the central fixation models predicted 32\%, while the PCA baseline models predicted 3\%.
We did not find any voxels that showed a significant difference in prediction between the gaze-aware model and the baseline model. 
This similarity in performance of baseline and gaze-aware encoding models is further supported by looking at the histograms of group average voxel performances for each model (see figure \ref{fig:fig1}a) and by looking at the distribution of voxels that are statistically significantly predicted by both models (see figure  \ref{fig:fig1}b). 
These figures show that both gaze-aware and baseline models significantly encode voxels in a range of visual stream areas from as early as V1 to as late as LO, FG, and STS.
These results suggest that gaze-aware encoding models match the results obtained by the traditional approach (which include information from the entire CNN feature set) with a fraction of the parameter space, and with added inter- and intra- individual variability.

When we look at differences between the two models (see Figure \ref{fig:fig1}c) in more detail, we find that the gaze-aware models perform better than baseline models in posterior occipital and parietal areas, while baseline models perform better in V1 and temporal areas. 
When we look at a set of predefined ROIs (see Figure \ref{fig:fig1}d), we observe no statistically significant differences between the models across participants.
However, performance for the baseline model is slightly higher than the gaze-aware model in V2 and STS. 
In addition to these regional differences, we observed that there was a lot of inter-individual variability in model performance, both for performance of the baseline model (see Supplementary Figure~\ref{fig:suppfig0}) , the gaze-aware model (see Supplementary Figure \ref{fig:suppfig1}) and the difference between the two models (see Supplementary Figures ~\ref{fig:suppfig2} and ~\ref{fig:suppfig4}).
We explore this variability across regions and participants more in the next section. 

Since we observed an advantage of baseline models in early visual areas that have a smaller population receptive field, we investigated whether including additional information about the voxel population receptive fields (pRF) in spatial feature selection could improve these models even more, especially for modelling early visual areas. 
We however did not observe such an advantage. 
The pRF adjusted gaze-aware models did not perform as well as the gaze only and baseline models (see Supplementary Figure~\ref{fig:suppfig5}). 
This might be due to the setup of our analysis pipeline, as we strongly downsampled the spatial features from the early VGG layers. 

\subsection{Where does the baseline model learn from? }
Given the large inter and intra-individual variability in performance differences between the baseline and the gaze-aware model, we investigated what might be driving these differences. 

We hypothesized that baseline models might perform better when they learn spatial weight distributions that are more similar to the gaze distribution. In these cases, gaze aware models may be less advantageous. 

\begin{figure*}[!ht]
\begin{center}
\includegraphics[scale=0.75,width=0.75\textwidth]{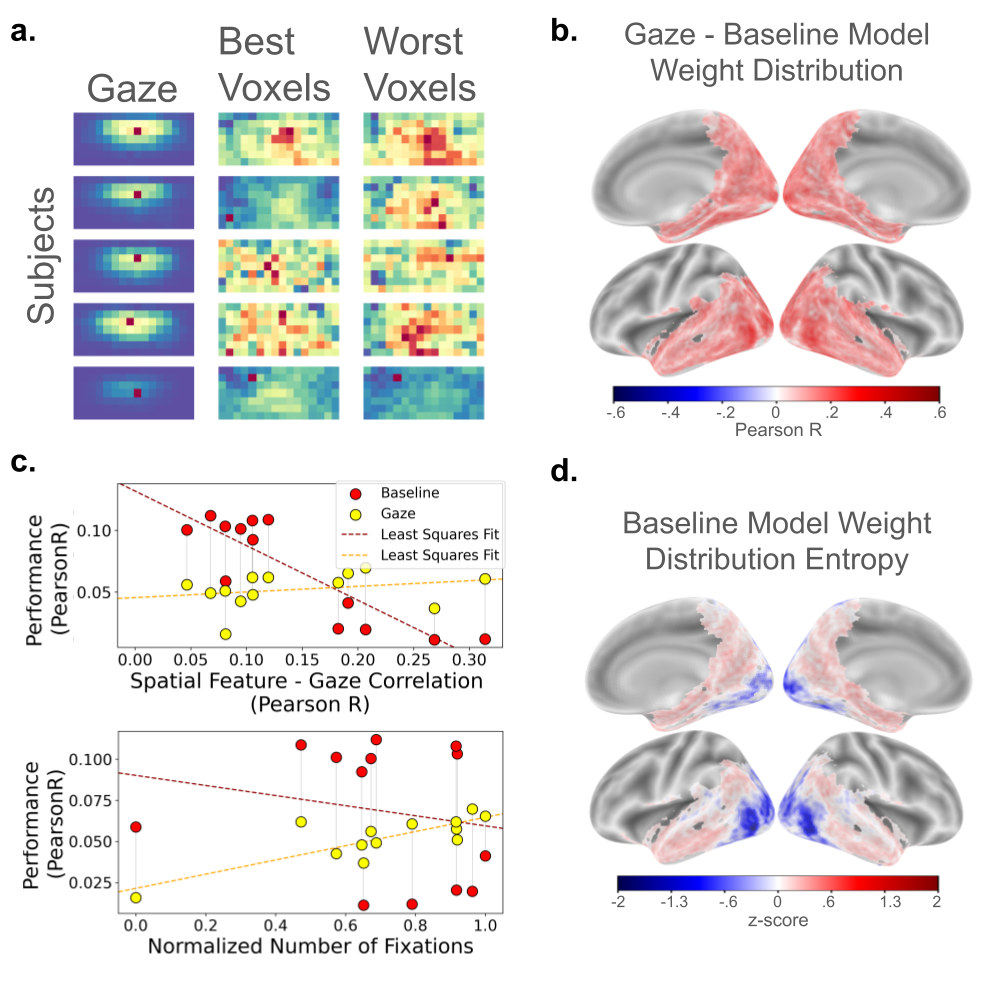}
\end{center}
\caption{\textbf{Baseline models learn from a spatial distribution that has some correspondence to the subject gaze, but only gaze-aware models improve with dynamic viewing.}
    (\textbf{a}) Each row shows one subject.
    Left column shows each subjects gaze distribution heatmap.
    Middle and Right columns shows the spatial distribution of weights learned by the baseline model.
    Middle shows the distribution of the best performing voxel.
    Right shows the distribution of the worst performing voxel.
    (\textbf{b}) Cortical maps show the group-averaged correlations between gaze distribution and each voxels spatial weight distributions learned by the baseline model. 
    (\textbf{c}) Top: Scatter plot shows the relationship between the model performances and similarity between subject gaze and learned spatial model weight distributions. Notice that the spatial weight distribution is a property of the baseline models. Therefore the x-axis for the gaze-aware models show the similarity between gaze distribution and the spatial weight distribution of the baseline model for the corresponding subject.
    Bottom: Scatter plot shows the relationship between the model performances and normalized number of fixations subjects made during free-viewing.
    (\textbf{d}) Shannon entropy of the spatial weight distributions learned by the baseline model, group-averaged and projected onto the cortex.
    } 
\label{fig:fig2}
\end{figure*}

In order to quantify the importance of each spatial location for the baseline model encoding, we computed the L2 norm (Euclidean norm) of the weight vectors from each spatial location (i.e. L2 norm computed over the channel dimension, while leaving the 7 by 16 spatial dimensions untouched).
Furthermore, to match the spatial dimensionality of the feature maps, we collected the pixel-resolution gaze distribution and downsampled it to a corresponding 7 by 16 grid.
Figure ~\ref{fig:fig2}a shows examples of the gaze distribution (left), best performing voxels spatial weight distribution (middle) and that of the worst performing voxel (right) for five random subjects. 
These plots are representative of the general pattern across participants, also in individual voxels. 
They show that the spatial weights learned by the baseline model are much more spread out than the gaze distribution.
Further, the peak of the spatial weights does not always correspond to the primary gaze location.  

Interestingly, the similarity between the gaze and the spatial weights was higher in lateral visual areas than in other regions (see Figure \ref{fig:fig2}b). 
These same areas also showed lower entropy of their weight distributions, suggesting that baseline models learn a less-distributed spatial weighting for these areas (see figure \ref{fig:fig2}d). 
You can see single subject maps of gaze and spatial weight similarities in Supplementary Figure~\ref{fig:suppfig3}.
This suggests that the baseline models do learn from more specific spatial locations in regions with smaller pRFs. 
However, all brain areas showed small to medium sized correlations between the gaze and spatial weight distributions, suggesting that the baseline model did not only learn from gaze locations, but other spatial locations were also driving encoding. 
Overall, the relationship between the correlation between subjects gaze distribution and voxels model spatial weight distributions and baseline model performances was weak but significant (average $r=0.07$), with p values ranging from $p<0.001$ to $p=0.008$ across subjects.

In addition to regional variability in performance, we also observed a great deal of variability between participants in model performance (see Supplementary Figures \ref{fig:suppfig0},\ref{fig:suppfig1} ~\ref{fig:suppfig2} and ~\ref{fig:suppfig4}). 
We expected that this may also be related to the alignment between the spatial weight distribution of the baseline model and the gaze distribution (see  Supplementary Figures \ref{fig:suppfig3}).
In other words, how much the baseline model learns from gaze locations.
Therefore, we computed the association between the mean performance of each participant (across all voxels) and the overall mean correlation between each subjects gaze and the spatial weight distributions (see Figure \ref{fig:fig2}c, top).
Contrary to our expectations, we can see that the baseline model performs better for subjects with low overall correlations between gaze and model spatial weight distributions, and worse for subjects with higher overall gaze-spatial weight distribution correlations (Pearson $r = -0.87$, $p <  0.001$).
In other words, subjects for whom the model weights better reflected the gaze showed poorer performance.
For the gaze-aware models, the same association is not present (Pearson $r = 0.26$, $p = 0.38$).
This unexpected effect suggests that the baseline model extracts gaze-correlated features only when encoding the most challenging of subjects' data.
For more straightforward subject data, the baseline model may be learning effectively from different parts of the image due to the naturally occurring correlation between different spatial features. 
Especially for subjects that have higher signal-to-noise neural data, the baseline model may learn well, even by learning from features that the subject did not look at directly.
Perhaps for subjects with more noisy data, the model is only able to stably read-out features located at positions that are aligned with the gaze location. 

To further explore the relationship between inter-individual differences in the gaze of participants and model performance, we also computed the relationships between gaze-aware and baseline models and number of fixations subjects made throughout the movie (see Figure \ref{fig:fig2}c bottom).
There is a strong correlation (Pearson $r = 0.81$, $p < 0.001$) between the number of fixations participants made and the performance of the gaze-aware models. 
This indicates that the dynamic feature selection inherent to gaze-aware models is indeed effective, and more so the more dynamic a subjects free-viewing is.
The relationship between the baseline model performance and number of fixations is much weaker, and not significant (Pearson $r = -0.19$, $p = 0.51$).

\section{Discussion}

We have demonstrated that gaze-aware brain encoding models, which leverage CNN features accounting for spatial characteristics derived from eye movements, perform as well as traditional spatially agnostic models in this dataset.
By reducing the parameter space in a spatially informed way, gaze-aware models are able to achieve interpretable and meaningful performance with fewer computational requirements. 
Moreover, incorporating naturalistic behaviour, such as eye-tracking, into the model opens new avenues for studying the brain under ecologically valid scenarios, providing deeper insights into how visual processing works in real-world contexts.

\subsection{Gaze-aware models compared to conventional models  }
The comparisons between our gaze-aware model and the baseline model reveal that models show significant encoding capacity across large parts of the ventral visual stream, despite the fact that participants were moving their eyes during movie viewing. 
Comparing these performances to those of encoding models in the literature is not straightforward due to variations in data collection (fMRI recording parameters, static vs. video stimuli, fixation vs. naturalistic) and analysis methods (preprocessing, linear response modelling).
Importantly, the performances we observe are on a different scale than most previous studies, partially because a noise ceiling, which was difficult to estimate due to the nature of the data, is not considered in this work.
Nonetheless, the statistically significantly predicted areas across the visual stream by gaze-aware models are very similar to studies which either use large (100+) group-averaged brain responses, or at least five times more scanning hours; most of which require fixation \citep{guclu_deep_2015, wen_neural_2018, han_variational_2019,khosla_cortical_2021,Kamali2025}. 
With this work we showed that, across all subjects, gaze-aware models achieve statistically significant encoding accuracies with only two hours of fMRI data and a fraction of the parameter space, explaining a similar breadth of activity in the ventral visual stream as reported in existing literature.

Our results show that conventional models learn from a spatial distribution that is broader than participants' actual gaze distributions, and that the spatial weight distributions do not always correspond to gaze location distributions.
This raises questions about the biological correspondence of what these models learn, particularly when participants are not fixating.
Previous work with fixed-gaze paradigms have successfully learned receptive field-like encoding structure \citep{StYves2018,Wang2019,Le2025}, but our findings suggest that when participants are free-viewing, models also learn from spatial locations that participants did not directly attend to.


The finding that baseline model weights are more spatially distributed than the gaze distribution suggests that non-fixated regions do contribute to voxel-wise encoding, though it is difficult to determine whether this reflects genuine peripheral contributions to voxel activity or the model's exploitation of natural spatial correlations within the feature set.
Peripheral and parafoveally attended content is known to support scene gist perception and to guide upcoming saccades, and regions outside the fixation point continue to drive responses in early visual areas through their retinotopic projections regardless of attentional allocation \citep{olivaBuildingGistScene2006,larsonContributionsCentralPeripheral2009,hendersonHumanGazeControl2003,torralbaContextualGuidanceEye2006}.
In higher areas such as STS and fusiform gyrus, where receptive fields integrate over broad swaths of the visual field, peripheral contributions may be proportionally greater, consistent with the modest baseline advantage we observe in those regions.
The gaze-aware model does not capture this signal, which likely accounts for part of the performance gap in those areas.


Importantly, we found that gaze-aware models were particularly beneficial for participants with more dynamic eye-movement patterns.
This is because the baseline model learns a fixed weight for each spatial location, while the gaze-aware model allows us for a dynamic weighting of different spatial location depending on the gaze. 
The strong positive correlation between the number of fixations and gaze-aware model performance demonstrates that this dynamic feature selection of our gaze-aware models becomes increasingly valuable as viewing behaviour becomes more active and exploratory.
This points to the specific advantages of gaze-aware modelling and that they become more pronounced in naturalistic settings where participants can actively navigate and interact with their environments.

\subsection{Task-free naturalistic vision}
While experimental control is important in cognitive neuroscience research, it is also important to recognize that the brain operates in a complex and dynamic environment.
Complete control over the data often comes with simplifications on our conceptual as well as technical models.
Brain encoding models that capture behaviour, rather than require it be constrained, offer a powerful tool for studying the brain's visual processing under naturalistic conditions.

Our approach represents a meaningful step toward more ecologically valid experimental paradigms.
The finding that our gaze-aware models were especially beneficial for participants who showed more dynamic eye movements during movie watching suggests that incorporating natural viewing behaviour provides genuine benefits for understanding naturalistic visual processing.
This is particularly relevant as the field moves towards paradigms involving gaming environments, virtual reality, and other interactive scenarios where constraining eye movements would fundamentally alter the nature of the tasks \citep{atari2020,Zhang2021,boyle2020courtois}.
In these paradigms, gaze-aware models are necessary, as fixation would fundamentally distort behaviour.
The reduction in computational demands achieved by this approach also has practical implications for the field.
The 112-fold reduction in model parameters makes encoding models more efficient to train, and may make encoding approaches more accessible to laboratories with limited resources.

\section{Limitations \& Future Directions}


In this paper, we drew a straightforward comparison between the conventional approach to brain-encoding models and our gaze-aware models in naturalistc settings.
In doing so, a number of factors were left unexplored which deserve further consideration.

Firstly, although naturalistic movie-viewing permits free eye movements and richer continuous stimulation than fixation-based paradigms, it remains a passive experience as viewers have no agency over the visual scene, and gaze is partly structured by professional cinematography rather than the self-generated exploratory behaviour that characterises truly active natural vision.
The gaze-aware framework introduced here is nonetheless well-suited to scale toward such active paradigms, precisely because it treats eye movements as a model input rather than a variable to be controlled.

Secondly, to ensure that the use of gaze information was straightforward, we modified the spatial resolution of the features from all layers of our CNN model such that all layers had an equivalent spatial resolution.
This may have led to a poorer encoder performance in early visual areas where spatial selectivity is highest and where our downsampling was greatest.
This may also explain why we found no additional benefit of adding information about the voxel pRFs into the model.

Thirdly, the conceptual simplicity and computational ease brought to the models by only using fixation events in our eye-tracking pre-processing might have hindered our models final performance.
Specifically, studies show that saccades and slow pursuits also elicit reliable responses in the early visual cortex \citep{Agtzidis2020}, two annotations which were ignored in this work.

Fourthly, the single dataset used in this study (two hours of fMRI data) and the absence of repeated stimuli might have constrained the model's ability to achieve higher accuracy.
As mentioned above, this also meant that a determination of a noise ceiling was not possible.
Future use of datasets which include repetition of stimuli within a free-viewing condition could prove useful.
Having said this, additional considerations must be made in such circumstances since subject gaze cannot be forceably controlled or repeated in the same manner.

Last but not least, the encoding performance of our models is limited by features extracted from the CNN model we selected.
The tasks and data which CNN models are trained on have the significant influence on the capabilities of any encoding model.
The CNN model used here was pre-trained on the ImageNet competition dataset \citep{imagenet}, which lacks ecological validity and does not optimally align with the statistical characteristics of the brain's natural visual environment \citep{mehrer2021ecoset}. 


In the future, we envision a general shift towards more precise and personalized encoding models that take inter- and intra-individual variances into account.
Mainstreaming this approach could not only reduce time and money spent on data collection but also, combined with promising new advancements like gaze prediction from fMRI recordings of the eyes \citep{frey_magnetic_2021}, a movement towards paradigms in which more naturalistic behaviour is embraced has potential for greater scientific insight. 
Gaze-aware models might be particularly beneficial for applications where efficiently training models with little data is imperative and individual behaviour is important to consider.
Examples include online perceptual reconstruction or visually driven neurofeedback setups, especially combined with fMRI data driven eye-tracking. 
Additionally, the targeted reduction in the dimensionality of CNN features, combined with the naturalistic and continuous brain data from movie watching could be further extended to recurrent neural networks which can capture time dependent influences of visual stimuli on brain activity.


Our current approach samples CNN features from a single point at the fixation coordinate, which leaves peripheral and parafoveal processing unmodelled.
A natural extension would be to sample a spatial kernel around the fixation point (for example a Gaussian-weighted region whose size scales with eccentricity) to better reflect the graded fall-off of spatial resolution and attention across the visual field.
Such an approach could recover some of the peripheral signal that the point sampling model misses, particularly for higher visual areas with large receptive fields, while retaining the parameter efficiency advantage over the full-feature baseline.
The size and shape of the kernel could itself be informed by each voxel's population receptive field, providing a principled link between the encoding model and retinotopic organisation.
This would also better reflect the known properties of visual attention, encompassing a broader attentional coverage that shifts continuously during naturalistic viewing \citep{davidNaturalStimulusStatistics2004,spillmannClassicalReceptiveField2015}.

More broadly, future work could ask targeted questions about the distinct roles of foveated and peripheral visual processing, the dynamics of covert and overt attention during naturalistic viewing, and the influence of multimodal cues on spatial attention allocation.
These questions would benefit from experimental designs purpose-built for those contrasts.


\section{Conclusion}
In this work we demonstrate that by building gaze-aware brain encoding models, we can build models with far fewer parameters while maintaining encoding performance and improving ecological validity over conventional brain encoding models.
This dramatic reduction in parameters needed for encoding models may facilitate more robust models as well as models which require far less data for training.
Importantly, our work also demonstrates that the performance of gaze-aware models, unlike baseline models, increases with increases in subject eye movements.
This provides evidence that for dynamic viewing conditions in which behaviour and perception are tightly coupled, such models may prove indispensable.
Finally, we showcased how trading-off experimental control with measurement of individual variations can bring us closer to a more ecologically valid investigation of brains.

\section*{Author Contributions}
\textbf{Dora Gözükara}: Conceptualization, Methodology, Software, Formal analysis, Visualization, Writing - Original Draft. 
\textbf{Nasir Ahmad}: Writing - Mathematical formulation \& Review \& Editing. 
\textbf{Katja Seeliger}: Writing - Review \& Editing. 
\textbf{Djamari Oetringer}: Methodology, Software, Writing - Review \& Editing. 
\textbf{Linda Geerligs}: Conceptualization, Methodology, Resources, Writing - Review \& Editing, Supervision.

\section*{Funding}
Linda Geerligs was supported by a Vidi grant (VI.Vidi.201.150) from the Netherlands Organization for Scientific Research.
Katja Seeliger was supported by a Max Planck Research Group grant to Martin N. Hebart. 

\section*{acknowledgements}
We thank Umut Güçlü for the initial conceptualisation and guidance, and Jordy Thielen for his help with pRF mapping.

\bibliography{precision}

\section{Supplementary}
\subsection{Single Subject Results}
\renewcommand{\figurename}{Supplementary Figure}
\setcounter{figure}{0}

To illustrate the actual model performances instead of a group averaged approximation, we plotted the gaze-aware model performances (see Supplementary Figure~\ref{fig:suppfig0}) and baseline model performances (see Supplementary Figure~\ref{fig:suppfig1}), and their differences (see Supplementary Figure~\ref{fig:suppfig2}) of example individual participants separately on the cortex.
We observe that the best performing voxels vary considerably between individuals.
This suggests that the group-level statistical testing we performed may be inappropriate in describing the actual benefit or harm of the gaze-aware encoding models.
Supplementary Figure~\ref{fig:suppfig4} shows the same single subject level results, but as grouped by selected ROIs.

\begin{figure}[!h]
\begin{center}
\includegraphics[width=0.9\columnwidth]{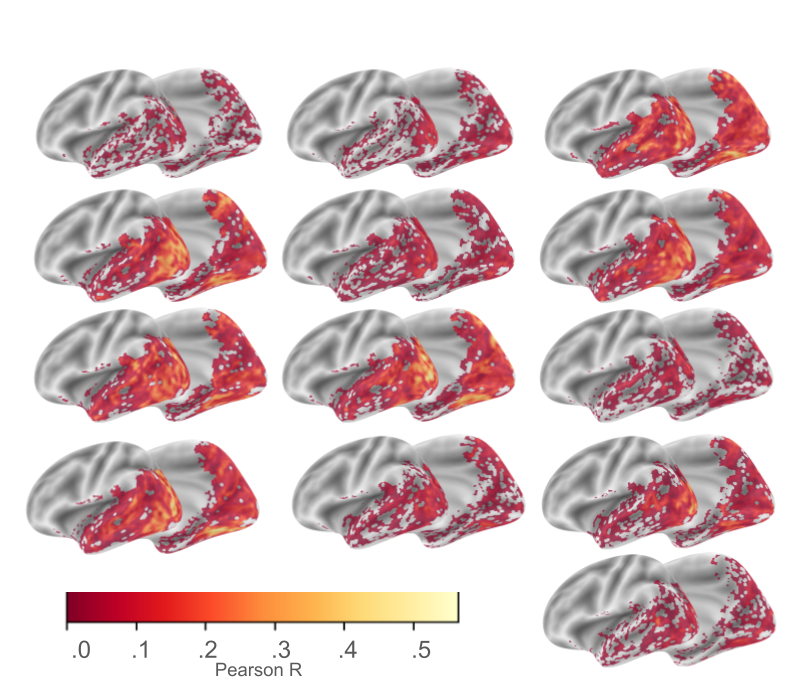}
\end{center}
\caption{\textbf{Single subject gaze-aware model performances.}
Cortical maps show gaze-aware model performances for all participants.
The best performing voxels vary considerably between individuals.
} 
\label{fig:suppfig0}
\end{figure}

\begin{figure}[!h]
\begin{center}
\includegraphics[width=0.9\columnwidth]{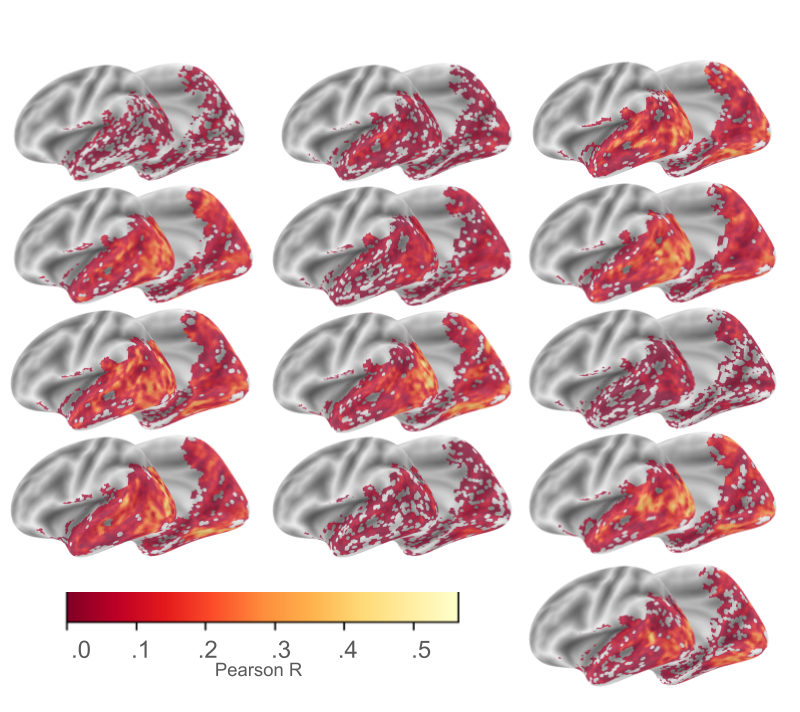}
\end{center}
\caption{\textbf{Single subject baseline model performances.}
Cortical maps show baseline model performances for all participants.
The best performing voxels vary considerably between individuals.
} 
\label{fig:suppfig1}
\end{figure}

\begin{figure}[!h]
\begin{center}
\includegraphics[width=0.9\columnwidth]{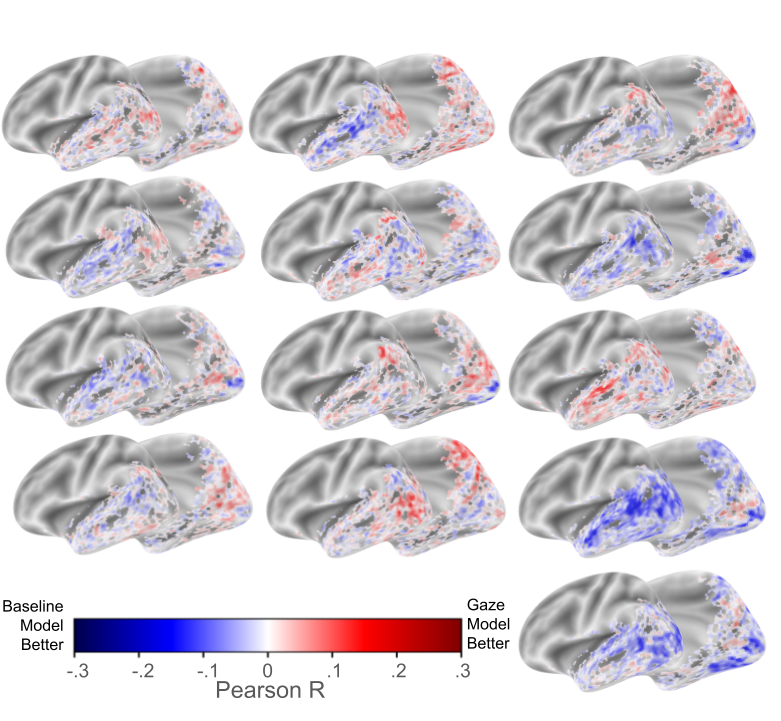}
\end{center}
\caption{\textbf{Single subject gaze-aware model baseline model differences.}
Cortical maps show baseline model performances for all participants.
The best performing voxels vary considerably between individuals.
Positive difference (red) means gaze-aware model is better.
} 
\label{fig:suppfig2}
\end{figure}

\begin{figure}[!h]
\begin{center}
\includegraphics[width=0.9\columnwidth]{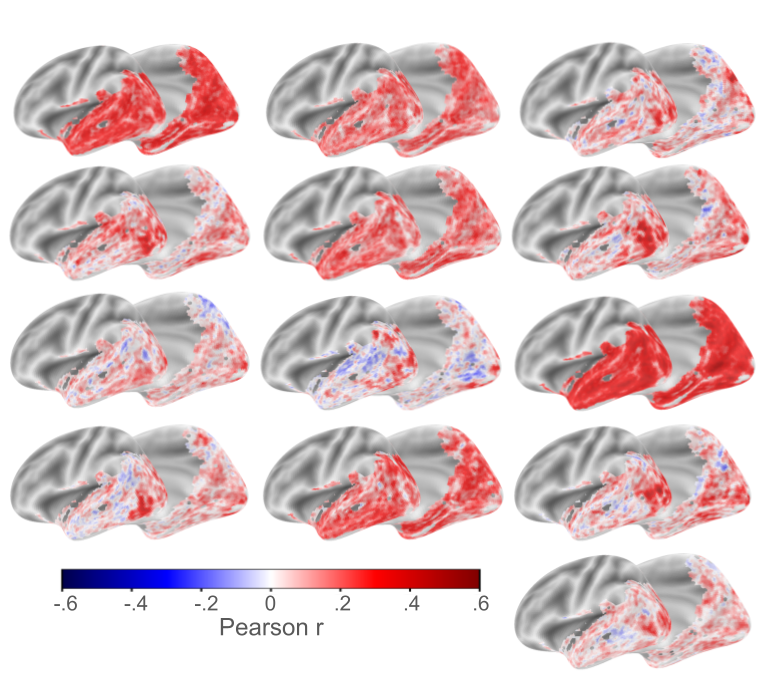}
\end{center}
\caption{\textbf{Single subject correlations between gaze distribution and the spatial distribution of weight norms learned by the baseline model for each voxel.}
} 
\label{fig:suppfig3}
\end{figure}

\begin{figure*}[!h]
\begin{center}
\includegraphics[width=\textwidth]{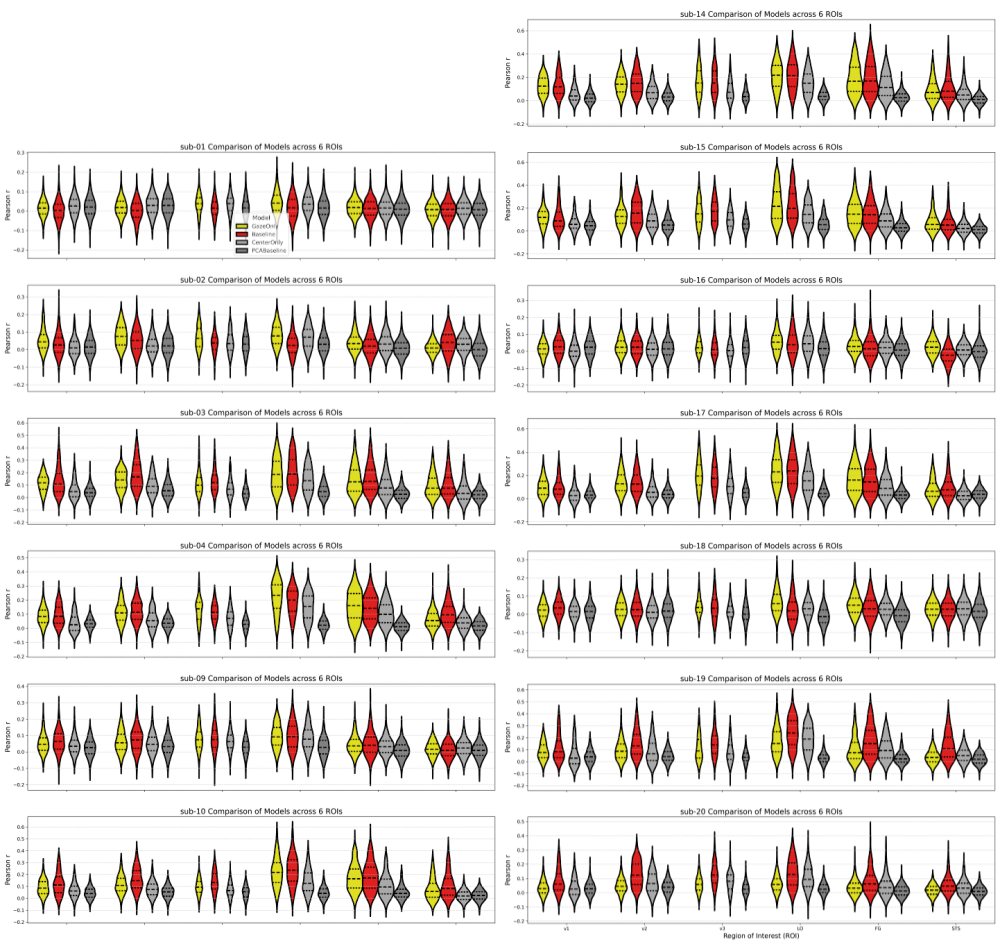}
\end{center}
\caption{\textbf{Single subject model performances vary across participants.}
Looking at single subject results within ROIs, it can be seen that some subjects and some ROIs are better captured by the gaze-aware model.
Each row shows within ROI voxel performances for a single subject.
Violin plot setups are same as Figure~\ref{fig:fig1}.
Lines within violins show quartiles.
    } 
\label{fig:suppfig4}
\end{figure*}

\subsection{Incorporating pRFs To Our Model}
\begin{figure*}[!h]
\begin{center}
\includegraphics[width=0.75\textwidth]{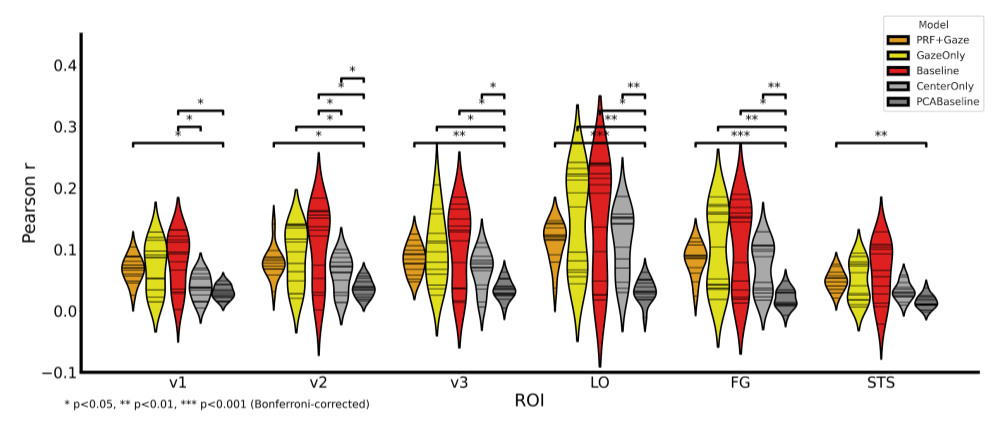}
\end{center}
\caption{\textbf{Incorporating pRF estimates did not improve gaze-aware model performance.}
Violin plots show model performances in bilateral V1, V2, V3, LOc, and FG.
Each violin plot shows the distribution of average model performance within an ROI across all participants.
Each horizontal line shows one participant.} 
\label{fig:suppfig5}
\end{figure*}

We've also trained gaze-aware models which incorporate voxel population receptive fields (pRF).
These models were created by off-setting the gaze coordinates for each subject with each of their specific voxels pRF estimate.
Below, we describe how these were obtained. 
Supplementary Figure~\ref{fig:suppfig5} shows how pRF off-setted gaze-aware models compare to other models.
Overall, pRF adjusted gaze-aware models did not perform as well as the gaze only and baseline models.
A few factors might have contributed to this rather counter-intuitive result.
First, voxel pRFs estimated with rotating wedges and expanding circles might not be generalising to naturalistic and dynamic vision.
Indeed, there is research which shows that pRFs are affected by factors like attention, which would dynamically change throughout movie-watching \citep{vanEs2018,Klein2014}.
Second, the spatial downsampling we applied to the earlier layers might have distorted the and/or prevented the potential benefits of including pRF information.

\subsubsection{Retinotopy Runs}
To estimate the receptive fields of different voxels, we used the retinotopic mapping data that is included in the StudyForrest dataset.
In these retinotopic mapping data, travelling wave stimuli were used to encode visual field representations in the brain via temporal activation patterns \citep{sengupta_studyforrest_2016}.
Expanding/contracting rings and clockwise/counter-clockwise wedges consisting of flickering radial chequerboards (flickering frequency of 5 Hz) were displayed on a grey background to map eccentricity and polar angle.
To estimate receptive fields for each voxel we use the ‘coarse to fine’ method described by \citet{dumoulin_population_2008}, as it is one of the most frequently used methods and is also computationally convenient.
This method estimates the most likely horizontal ($x$) coordinate, vertical ($y$) coordinate, and size of a voxels pRF based upon an iterative process, starting with a coarse estimate and proceeding to a finer one.
We used the $x$ and $y$ coordinates of the fine estimate for each voxel.
As our models also covered brain areas outside of the immediate visual stream, we did not exclude voxels using an explained variance threshold of each voxel’s pRF model.
The fine to coarse method estimates $x$ and $y$ coordinates in visual degrees in the range $[-10, 10]$.
Based on the distance of the screen from the participants eyes (65 cm), we convert the visual degrees to pixel coordinates on the screen.
The pRFs were estimated based on the native subject space images that were aligned with the movie-viewing-runs. After estimation, the $x$ and $y$ coordinates and the pRF size were transformed to MNI space. 






\end{document}